\documentclass[11pt,a4paper]{article}
\usepackage{amsmath,latexsym,fullpage,amssymb,color,couleurs}
\usepackage{ifthen,graphics,epsfig}
\usepackage[english]{babel}
\usepackage{times}
\usepackage{float}
\usepackage{xspace}
\usepackage[T1]{fontenc}
\newlength{\squarewidth}

\newtheorem{theorem}{Theorem}
\newtheorem{lemma}{Lemma}

\newcommand{\toto}{xxx}
\newenvironment{proofT}{\noindent{\bf Proof }}
{\hspace*{\fill}$\Box_{Theorem~\ref{\toto}}$\par\vspace{3mm}}
\newenvironment{proofL}{\noindent{\bf Proof }}
{\hspace*{\fill}$\Box_{Lemma~\ref{\toto}}$\par\vspace{3mm}}

\newenvironment{lemma-repeat}[1]{\begin{trivlist}
\item[\hspace{\labelsep}{\bf\noindent Lemma~\ref{#1} }]}%
{\end{trivlist}}

\newenvironment{theorem-repeat}[1]{\begin{trivlist}
\item[\hspace{\labelsep}{\bf\noindent Theorem~\ref{#1} }]}%
{\end{trivlist}}

\newcounter{linecounter}
\newcommand{\linenumbering}{\ifthenelse{\value{linecounter}<10}
{(\arabic{linecounter})}{(\arabic{linecounter})}}
\renewcommand{\line}[1]{\refstepcounter{linecounter}\label{#1}\linenumbering}
\newcommand{\resetline}[1]{\setcounter{linecounter}{0}#1}
\renewcommand{\thelinecounter}{\ifnum \value{linecounter} > 
9 \else \fi\arabic{linecounter}}
\newfloat{algorithm}{thp}{lop}
\floatname{algorithm}{Algorithm}
\newfloat{construction}{thp}{lop}
\floatname{construction}{Construction}
\newcommand{\Xomit}[1]{}
\newcommand{\REG}{\mathit{REG}}
\newcommand{\CAMP}{{\cal CAMP}}

\begin{document}


\title{\bf Time-Efficient Read/Write  Register\\
           in Crash-prone Asynchronous Message-Passing Systems}

\author{Achour Most\'efaoui$^{\dag}$, 
        Michel Raynal$^{\star,\ddag}$\\~\\
$^{\dag}$LINA, Universit\'e de Nantes, 44322 Nantes, France \\
$^{\star}$Institut Universitaire de France\\
$^{\ddag}$IRISA, Universit\'e de Rennes, 35042 Rennes, France \\
~\\~\\ Tech Report \#2031, 14 pages, January 2016\\ 
IRISA, University of Rennes 1, France
}

\date{}

\maketitle


\begin{abstract}
The atomic register is certainly the most basic object of computing
science.  Its implementation on top of an $n$-process asynchronous
message-passing system has received a lot of attention.  It has been
shown that $t<n/2$ (where $t$ is the maximal number of processes that
may crash) is a necessary and sufficient requirement to build an
atomic register on top of a crash-prone asynchronous message-passing system.
Considering such a context, this paper visits the notion of a fast
implementation of an atomic register, and presents a new time-efficient
asynchronous algorithm. Its time-efficiency is measured  according to two 
different underlying synchrony assumptions. Whatever this assumption, 
a write operation always costs a round-trip delay, while a 
read operation costs always a round-trip delay in favorable circumstances
(intuitively, when it is not concurrent with a write). 
When designing this algorithm, the design spirit was to be  as close as 
possible to the one of the famous ABD algorithm (proposed by Attiya, Bar-Noy,
and Dolev).
~
\end{abstract}

\newpage

\section{Introduction}
Since Sumer time~\cite{K56}, and --much later-- Turing's machine
tape~\cite{T36}, read/write objects are certainly the most basic
memory-based communication objects.  Such an object, usually called a
{\it register}, provides its users (processes) with a write operation 
which defines the new value of the register, and a read operation which 
returns the value of the register. 
When considering sequential computing, registers are universal in the sense 
that they allow to solve any problem that can be solved~\cite{T36}.

\paragraph{Register in message-passing systems}
In a message-passing system, the computing entities communicate only
by sending and receiving messages transmitted through a communication network.
Hence, in such a system, a register is not a communication object
given for free, but constitutes a communication abstraction which must
be built with the help of the communication network and the local memories of
the processes.

Several types of registers can be defined according to which processes
are allowed to read or write it, and the quality (semantics) of the
value returned by each read operation.  We consider here registers
which are single-writer multi-reader (SWMR), and atomic. 
Atomicity means that (a) each read or write operation appears
as if it had been  executed instantaneously at a single point of the time
line, between is start event and its end event, 
(b) no two operations appear at the same point of the time line, and 
(c) a read returns the value written
by the closest preceding write operation (or the initial value of the
register if there is no preceding write)~\cite{L86}.  Algorithms building
multi-writer multi-reader (MWMR) atomic registers from single-writer
single-reader (SWSR) registers with a weaker semantics (safe or regular
registers) are described in several textbooks (e.g.,~\cite{AW04,L96,R13}).

Many distributed algorithms have been proposed, which build a
register on top of a message-passing system, be it failure-free or
failure-prone. In the failure-prone case,  the addressed failure models 
are the process crash failure model, or the  Byzantine process failure 
model (see, the textbooks~\cite{AW04,L96,R10,R13-a}).  The most famous of 
these algorithms was proposed by H. Attiya, A. Bar-Noy,  and D. Dolev 
in~\cite{ABD95}.  This algorithm, which is usually called ABD
according to the names its authors, considers an $n$-process asynchronous 
system in which up to $t<n/2$ processes may crash (it is also shown 
in~\cite{ABD95} that  $t<n/2$ is an upper bound 
of the number of process crashes which can be tolerated). 
This simple and elegant algorithm, 
relies on (a) quorums~\cite{V12}, and (b) a simple broadcast/reply 
communication pattern. ABD uses this pattern once in  a write operation, 
and twice in a read operation implementing an SWMR register.

\paragraph{Fast operation} 
To our knowledge, the notion of a {\it fast implementation} of an  atomic
register operation,  in failure-prone asynchronous message-passing 
systems, was introduced  in~\cite{DGLC04} for process crash
failures, and in~\cite{DGLV10} for Byzantine process failures.  These
papers consider a three-component model, namely there are three different 
types of processes: a set of writers $W$, a set of readers $R$, and a set of
servers $S$ which implements the register. 
Moreover, a client (a writer or a reader) can communicate
only with the servers, and the servers do not communicate among themselves.

In these papers, {\it fast} means that a read or write operation must
entail exactly one communication round-trip delay between a client (the
writer or a reader) and the servers. When considering the process
crash failure model (the one we are interested in in this paper), it is
shown in~\cite{DGLC04} that, when $(|W|=1)\wedge(t\geq 1)\wedge(|R|\geq 2)$, 
the condition $(|R|< \frac{|S|}{t}-2)$ is necessary and sufficient to have
fast read and write operations (as defined above), which implement an 
atomic register. It is also shown in~\cite{DGLC04} that
there is no fast implementation of an MWMR atomic register if
$\big((|W|\geq 2)\wedge (|R|\geq 2)\wedge (t\geq 1)\big)$.

\paragraph{Content of the paper}
The work described in~\cite{DGLC04,DGLV10} is mainly on the limits of
the three-component model (writers, readers, and servers constitute
three independent sets of processes) in the presence of process crash
failures, or Byzantine process failures. These limits are captured by
predicates involving the set of writers ($W$), the set of readers
($R$), the set of servers ($S)$, and the maximal number of servers
that can be faulty ($t$).  Both the underlying model used in this
paper and its aim are different from this previous work.

While keeping the spirit (basic principles and simplicity) of ABD, our
aim is to design a {\it time-efficient} implementation of an atomic 
register in the classical model  used in many articles and textbooks
(see, e.g.,~\cite{ABD95,AW04,L96,R13}).  This model,
where any process can communicate with any process, can be seen as a
peer-to-peer model in which each process is both a client (it can
invoke operations) and a server (it manages a local copy of the
register that is built).\footnote{Considering the three-component model 
where each reader is also a server (i.e., $R=S$), we obtain a two-component 
model with one  writer and reader-server processes. In this model, the
necessary  and sufficient condition $(|R|< \frac{|S|}{t}-2)$ can never be
satisfied, which means that, it is impossible to design a fast 
implementation of a SWMR atomic register in such a two-component model.}

Adopting the usual distributed computing assumption that 
(a) local processing times are negligible and assumed consequently to have  
zero duration, and (b) only communication takes time, this paper focuses
on the communication time needed to complete a read or write operation.  
For this reason the term {\it time-efficiency} is defined here in
terms on message transfer delays, namely, the cost of a read or write
operation is measured by the number of ``consecutive'' message transfer 
delays they require to terminate. Let us notice that this
includes transfer delays due to causally related messages (for example
round trip delays generated by request/acknowledgment messages), but
also (as we will see in the proposed algorithm) message transfer delays
which occur sequentially without being necessarily causally related. 
Let us notice that this notion of a time-efficient operation does not 
involve the model parameter $t$.

In order to give a precise meaning to the notion of a ``time-efficient
implementation'' of a register operation, this paper considers two distinct
ways to measure the duration of read and write operations, each based
on a specific additional synchrony assumption.  One is the ``bounded
delay'' assumption, the other one the ``round-based synchrony''
assumption.  More precisely, these assumptions and the associated
time-efficiency of the proposed algorithm are the following.
\begin{itemize}
\vspace{-0.1cm}
\item {\it Bounded delay}  assumption. \\
Let us assume that every message 
takes at most $\Delta$ time units to be transmitted from its sender to any
of its receivers. In such a context, the algorithm presented in the paper
has the following time-efficiency properties.
\begin{itemize}
\vspace{-0.2cm}
\item A write operation takes at most $2\Delta$ time units.
\vspace{-0.1cm}
\item 
A read operation which is  write-latency-free takes
at most $2\Delta$ time units. (The notion of write-latency-freedom is
defined in Section~\ref{sec:fast-definition}. Intuitively,  
it captures the fact that the behavior of the read does not depend
on a concurrent or faulty write operation, which is the usual case 
in read-dominated applications.)
Otherwise, it takes at most $3\Delta$ time units, except in the case where 
the read operation is concurrent with a write operation and the writer 
crashes during this write, where it can take up to $4\Delta$ time units.
(Let us remark that a process can experience at
most once the $4\Delta$ read operation scenario.)
\end{itemize} 
\vspace{-0.4cm}
\item {\it Round-based synchrony}  assumption. \\
Here, the underlying communication system is assumed to be 
round-based synchronous~\cite{AW04,L86,R13-a}. In such a system, 
the processes progress by executing consecutive synchronous rounds. 
In every round, according to its code, 
a process possibly sends a message to a subset of processes, 
then receives all the  messages sent to it during the current round, 
and finally executes local computation. At the end of a round,
all processes are directed to simultaneously progress to the next round. 
In such a synchronous system, everything appears as if  all messages take  
the very same time to go from their sender to theirs receivers, 
namely the duration  $\delta$ associated with a round. When executed in such a 
context,  the proposed algorithm has the following time-efficiency properties.
\begin{itemize}
\vspace{-0.2cm}
\item The duration of a write operation  is $2\delta$ time units.
\vspace{-0.1cm}
\item The duration of a read operation is $2\delta$ time units, 
except possibly in the specific scenario where the writer crashes while 
executing the write operation concurrently with the read, 
in which case the duration of the read can be $3\delta$ time units
(as previously, let us remark that a process can experience at
most once the $3\delta$ read operation scenario.)
\end{itemize}
\end{itemize}

Hence, while it  remains correct in the presence of 
any asynchronous message pattern (e.g., when each message takes 
one more time unit than any previous  message), the proposed algorithm 
is particularly time-efficient when  ``good'' scenarios  occur. Those
are the ones defined by  the previous synchrony patterns where the duration 
of a read or a write operation corresponds to  a single round-trip delay. 
Moreover, in the other synchronous scenarios, where a
read operation is concurrent with a write, the maximal duration of the
read operation is precisely quantified. A concurrent write 
adds uncertainty whose resolution by a read operation requires one more 
message transfer delay (two in the case of the $\Delta$ synchrony assumption, 
if the concurrent write crashes).

\paragraph{Roadmap}
The paper consists of~\ref{sec:conclusion} sections.
Section~\ref{sec:model-definitions} presents the system model.
Section~\ref{sec:fast-definition} defines the atomic register
abstraction, and the notion of a time-efficient implementation.  Then,
Section~\ref{sec:algorithm} presents an asynchronous algorithm providing 
an implementation of an atomic register with time-efficient operations, 
as previously defined. Section~\ref{sec:proof} proves its properties.
Finally, Section~\ref{sec:conclusion} concludes the paper.

\section{System Model}
\label{sec:model-definitions}

\paragraph{Processes}
The computing model is composed of a set of $n$ sequential processes
denoted $p_1$, ..., $p_n$. Each process is asynchronous which means
that it proceeds at its own speed, which can be arbitrary and remains
always unknown to the other processes.  

A process may halt prematurely (crash failure), but executes correctly
its local algorithm until it possibly crashes. The model parameter $t$
denotes the maximal number of processes that may crash in a  run.
A process that crashes in a run is said to be {\it faulty}. Otherwise, 
it is {\it correct} or {\it non-faulty}.

\paragraph{Communication}
The processes cooperate by sending and receiving messages through 
bi-directional channels. The communication network is a complete network, 
which means that any process $p_i$  can directly send a message to any 
process $p_j$ (including itself). 
Each channel is reliable (no loss, corruption, nor creation of messages),
not necessarily first-in/first-out, and asynchronous (while the transit time 
of each message is finite, there is no upper bound on message transit times).  

A process $p_i$ invokes the operation 
``${\sf send}$ {\sc tag}($m$) {\sf to} $p_j$'' to send  $p_j$ the message 
tagged {\sc tag} and carrying the value $m$. It receives a message 
tagged {\sc tag} by invoking the operation ``${\sf receive}$ {\sc tag}()''. 
The macro-operation ``${\sf broadcast}$ {\sc tag}($m$)'' is a 
shortcut for ``{\bf for each}  $j\in \{1,\ldots,n\}$  
${\sf send}$ {\sc tag}($m$) {\sf to} $p_j$ {\bf end for}''. 
(The sending  order is arbitrary, which means that, if the sender 
crashes while executing this statement, an arbitrary -- possibly empty-- 
subset of processes will receive the message.)

Let us notice that, due to process and message asynchrony, no process can
know if an other process crashed or is only very slow.

\paragraph{Notation} 
In the following, the previous computation model, restricted to 
the case where $t<n/2$, is denoted ${\CAMP}_{n,t}[t<n/2]$
(Crash Asynchronous Message-Passing).  

It is important to notice that, in this model, all processes are a priori
``equal''. As we will see, this allows each process to be at the same
time a ``client'' and a ``server''.  In this sense, and as noticed in the 
Introduction, this  model is the ``fully connected peer-to-peer'' model 
(whose structure is 
different from other computing models such as the  client/server model,
where  processes are partitioned into clients and  servers, 
playing different roles).

\section{Atomic Register and Time-efficient Implementation}
\label{sec:fast-definition}

\subsection{Atomic register}
A {\it concurrent object} is an object that can be accessed by several 
processes (possibly simultaneously).  An SWMR {\it atomic} 
register (say $\REG$) is a concurrent object which provides exactly one 
process (called the writer) with an operation denoted $\REG.{\sf write}()$, 
and all processes with an operation denoted $\REG.{\sf read}()$.  
When the writer invokes $\REG.{\sf write}(v)$ it defines
$v$ as being the new value of $\REG$.  An SWMR atomic register
(we also say the register is {\it linearizable}~\cite{HW90})  is
defined by the following set of properties~\cite{L86}.

\begin{itemize}
\vspace{-0.1cm}
\item Liveness. An invocation of an operation by a correct process terminates.
\vspace{-0.2cm}
\item Consistency (safety).  All the operations invoked by the
processes, except possibly --for each faulty process-- the last
operation it invoked, appear as if they have been executed
sequentially and this sequence of operations is such that:
\begin{itemize}
\vspace{-0.1cm}
\item each read returns the value written by the closest write that precedes
it (or the initial value of $\REG$ if there is no preceding write), 
\vspace{-0.1cm}
\item if an operation $op1$ terminated before an operation $op2$ started, then 
 $op1$ appears before $op2$ in the sequence.
\end{itemize}
\end{itemize}

This set of properties states that, from an external observer point of
view, the object appears as if it was accessed sequentially by the
processes, this sequence (a) respecting the real time access order,
and (ii) belonging to the sequential specification of a read/write
register. 

\subsection{Notion of a time-efficient  operation}
The notion of a time-efficient operation is not related to its correctness, 
but is a property of its implementation. It is sometimes called
{\it non-functional} property. In the present case, it captures the time
efficiency of operations.\footnote{Another example of a non-functional
  property is {\it quiescence}. This property is on  algorithms implementing
  reliable communication on top of unreliable  networks~\cite{ACT00}.
  It states that the number of  underlying implementation messages 
  generated by an application message must be finite. Hence, if 
  there is a time after which no application process sends messages, 
  there is a time after which the system is quiescent.}

As indicated in the introduction, we consider here two synchrony assumptions 
to define what we mean by time-efficient  operation implementation.
As we have seen, both are based on the duration 
of  read and write operations, in terms of message transfer delays.
Let us remember that, in both cases, it is assumed that the local processing 
times needed to implement these high level read and write operations are 
negligible. 

\subsubsection{{\it Bounded delay}-based definition of a time-efficient
 implementation}
Let us assume an underlying communication system where message
transfer delays are upper bounded by $\Delta$.  

\paragraph{Write-latency-free read operation and interfering write}
Intuitively, a read operation is {\it write-latency-free} if its execution 
does ``not interleave'' with the execution of a write operation. 
More precisely, let $\tau_r$ be the starting time of a read operation.
This read  operation  is {\it write-latency-free} if 
(a) it is not concurrent  with a write operation, and
(b) the closest preceding write did not crash and started at 
a time $\tau_w < \tau_r-\Delta$. 

Let ${\sf opr}$ be a read operation, which started at time $\tau_r$. 
Let ${\sf opw}$ be the closest write  preceding ${\sf opr}$.
If  ${\sf opw}$ started at time   $\tau_w \geq \tau_r-\Delta$, it is said to 
be {\it interfering} with ${\sf opr}$.

\paragraph{{\it Bounded delay}-based definition}
An implementation of a read/write register is {\it time-efficient} 
(from a bounded delay point of view) if it satisfies the following properties.
\begin{itemize}
\vspace{-0.1cm}
\item A write operation takes at most $2\Delta$ time units.
\vspace{-0.2cm}
\item A read operation which is write-latency-free
 takes  at most $2\Delta$ time units. 
\vspace{-0.2cm}
\item   A read operation which is not write-latency-free takes at most 
\begin{itemize}
\vspace{-0.2cm}
\item $3\Delta$ time units if the writer does not crash
while executing the interfering write, 
\vspace{-0.1cm}
\item $4\Delta$ time units if the writer crashes while executing the 
interfering write (this scenario can appear at most once for each process).
\end{itemize}
\end{itemize}

\subsubsection{{\it Round synchrony}-based definition of a time-efficient
 implementation}
Let us assume that the underlying communication system is round-based 
synchronous, where each message transfer delay is equal to $\delta$. 
When considering this underlying synchrony assumption, 
it is assumed that a process sends or broadcasts at most one message 
per round, and this is done at the beginning of a round. 

An implementation of a read/write register is {\it time-efficient} 
(from the round-based synchrony point of view) if it satisfies the 
following properties.
\begin{itemize}
\vspace{-0.1cm}
\item The duration of a write operation  is $2\delta$ time units.
\vspace{-0.2cm}
\item The duration of a read operation is $2\delta$ time units, 
except possibly in the ``at most once''  scenario  where the writer crashes 
while executing the write operation concurrently with the read, 
in which case the duration of the read can be $3\delta$ time units.
\end{itemize}

\paragraph{What does the proposed algorithm}
As we will see, the proposed algorithm, designed for the asynchronous system 
model $\CAMP_{n,t}[t<n/2]$, provides an  SWMR atomic register implementation 
which is time-efficient  for both its ``bounded delay''-based definition, 
and its ``round synchrony''-based  definition.

\section{An Algorithm with Time-efficient Operations}
\label{sec:algorithm}

The design of the algorithm, described in Figure~\ref{algo:fast-basic-version}, 
is voluntarily formulated  to be as close as possible to ABD. 
For the reader aware of ABD, this will help its understanding.

\paragraph{Local variables}
Each process $p_i$ manages the following local variables.
\begin{itemize}
\vspace{-0.1cm}
\item $reg_i$ contains the value of the constructed register $\REG$, 
as currently known by $p_i$. It is initialized to the initial value of  
$\REG$ (e.g., the default value $\bot$).
\vspace{-0.2cm}
\item $wsn_i$ is the sequence number associated with the value in $reg_i$.
\vspace{-0.2cm}
\item  $rsn_i$ is the sequence number of the last read operation 
invoked by $p_i$. 
\vspace{-0.2cm}
\item $swsn_i$ is a synchronization local variable. 
It contains the sequence number of the most recent value of $\REG$
that, to $p_i$'s knowledge, is known by at least $(n-t)$ processes. 
This variable (which is new with respect to other algorithms) 
is at the heart of the time-efficient implementation of the read operation. 
\vspace{-0.2cm}
\item $res_i$ is the value  of $\REG$ whose sequence number is $swsn_i$. 
\end{itemize}

\begin{figure}[th]
\centering{\fbox{
\begin{minipage}[t]{150mm}
\footnotesize
\renewcommand{\baselinestretch}{2.5}
\resetline
\begin{tabbing}
aaaaa\=aaa\=aaaaa\=aaaaaa\=\kill

{\bf local variables initialization:}
 $reg_i \gets \bot$; $wsn_i \gets 0$; $swsn_i \gets 0$; $rsn_i \gets 0$.\\~\\

{\bf operation} $\mathsf{write}$($v$) {\bf is} 
\\

\line{F-SWMR-01} 
\> $wsn_i\leftarrow wsn_i +1$; $reg_i\leftarrow v$;
   $\mathsf{broadcast}$ {\sc write}($wsn_i,v$);\\

\line{F-SWMR-02} 
\> {\bf wait} 
 \big({\sc write}($wsn_i,-$) received from $(n-t)$ different processes\big);\\
  
\line{F-SWMR-03} 
\>  ${\sf return}()$\\
{\bf end operation}.\\~\\

{\bf operation} $\mathsf{read}$() {\bf is}
\% the writer may directly return $reg_i$ \% \\

\line{F-SWMR-04} 
\> $rsn_i\gets rsn_i+1$; $\mathsf{broadcast}$ {\sc read}($rsn_i$);\\

\line{F-SWMR-05} 
\> {\bf wait} \= \big( (messages {\sc state}$(rsn,-)$ received from $(n-t)$ 
  different processes)  $\wedge$ ($swsn_i \geq maxwsn$) \\

\>\>  $~$ where $maxwsn$ is the greatest sequence number 
    in the previous {\sc state}$(rsn,-)$ messages\big);\\

\line{F-SWMR-06} 
\>  ${\sf return}(res_i)$\\
{\bf end operation}. \\
\%-----------------------------------------------------------------------------------------------------------------------\\~\\

{\bf when} {\sc write}($wsn,v$) {\bf is received do}\\

\line{F-SWMR-07} 
\> {\bf if} ($wsn>wsn_i$) 
   {\bf then}  $reg_i \leftarrow v$; $wsn_i\gets wsn$ {\bf end if}; \\

\line{F-SWMR-08} 
\> {\bf if} (not yet done) 
   {\bf then} $\mathsf{broadcast}$ {\sc write}($wsn,v$)
{\bf end if}; \\

\line{F-SWMR-09} 
\> {\bf if} \big({\sc write}($wsn,-$)   received 
                 from $(n-t)$ different processes\big)  \\

\line{F-SWMR-10} 
\> \> 
 {\bf then} {\bf if} $(wsn > swsn_i)$ $\wedge$ (not already done)  {\bf then}
             $swsn_i\gets wsn$;  $res_i\gets v $ {\bf end if}\\

\line{F-SWMR-11} 
\>   {\bf end if}. \\~\\

{\bf when} 
{\sc read}$(rsn)$ {\bf is received from} $p_j$ {\bf do}\\

\line{F-SWMR-12} 
\> $\mathsf{send}$ {\sc state}$(rsn,wsn_i)$ $\mathsf{to}$ $p_j$. 

\end{tabbing}
\end{minipage}
}
\caption{Time-efficient SWMR atomic register in ${\cal AMP}_{n,t}[t<n/2]$}
\label{algo:fast-basic-version}
}
\end{figure}

\paragraph{Client side:  operation ${\sf write}()$ invoked by the writer}
When the writer $p_i$ invokes $\REG.{\sf write}(v)$, 
it increases $wsn_i$, updates $reg_i$, and broadcasts the message 
{\sc write}$(wsn_i,v)$ (line~\ref{F-SWMR-01}).  Then, it 
waits until it has received an acknowledgment message 
from $(n-t)$ processes  (line~\ref{F-SWMR-02}). 
When this occurs, the operation terminates (line~\ref{F-SWMR-03}).
Let us notice that the acknowledgment message is a copy of 
 the very same message as the one it broadcast. 

\paragraph{Server side: reception of a message  ${\sc write}(wsn,v)$}
when a process $p_i$ receives such a message, and this message carries a 
more recent value than the one currently stored in $reg_i$,  
$p_i$ updates accordingly $wsn_i$ and $reg_i$
(line~\ref{F-SWMR-07}). Moreover, if this message is the first message
carrying the sequence number $wsn$,  $p_i$ 
forwards to all  the processes the message  {\sc write}$(wsn,v)$ it has 
received (line~\ref{F-SWMR-08}). 
This broadcast has two aims: to be an acknowledgment for the writer, and 
to inform the other processes  that  $p_i$ ``knows'' this value.\footnote{Let 
us observe that, due to asynchrony, it is possible  that $wsn_i>wsn$ 
when $p_i$ receives a message {\sc write}$(wsn,v)$ for the first time.}

Moreover, when $p_i$ has received the  message {\sc write}$(wsn,v)$ 
from $(n-t)$ different processes, and  $swsn_i$ is smaller than $wsn$, 
it  updates its local synchronization variable $swsn_i$ 
and accordingly assigns $v$ to $res_i$ (lines~\ref{F-SWMR-09}-\ref{F-SWMR-11}).

\paragraph{Server side: reception of a message  {\sc read}$(rsn)$}
When a process $p_i$ receives such a message from a process $p_j$, it
sends by return to $p_j$ the message {\sc state}$(rsn,wsn_i)$, thereby
informing it on the freshness of the last value of $\REG$ it knows
(line~\ref{F-SWMR-12}).  The parameter $rsn$ allows the sender $p_j$
to associate the messages {\sc state}$(rsn,-)$ it will receive with
the corresponding request identified by $rsn$.

\paragraph{Client side: operation ${\sf read}()$ invoked by a process $p_i$}
When a process invokes $\REG.{\sf read}()$, it first broadcasts the
message {\sc read}$(rsn_i)$ with a new sequence number.  Then, it
waits until ``some'' predicate is satisfied (line~\ref{F-SWMR-05}), and
finally returns the current value of $res_i$.  Let us notice that the
value $res_i$ that is returned is the one whose sequence number is $swsn_i$.

The waiting predicate is the heart of the algorithm. Its first part 
states that $p_i$ must have received a message {\sc  state}$(rsn,-)$ 
from $(n-t)$ processes. Its second part, namely $(swsn_i \geq maxwsn)$,  
states that the  value in $p_i$'s local variable $res_i$ is 
as recent or more recent than the value associated with the greatest write 
sequence number $wsn$ received by $p_i$ in a message {\sc state}$(rsn,-)$. 
Combined with the broadcast of messages {\sc write}$(wsn,-)$ 
issued by each process at line~\ref{F-SWMR-08}, this waiting 
predicate ensures both the correctness of the returned value (atomicity), 
and the fact that the read implementation is time-efficient.

\section{Proof of the Algorithm}
\label{sec:proof}

\subsection{Termination and atomicity}
The properties proved in this section are independent 
of the message transfer delays (provided  they are finite).

\begin{lemma}
\label{lemma:termination}
If the writer is correct, all its write invocations terminate.  
If a reader  is correct, all its read invocations terminate. 
\end{lemma}

\begin{proofL}
Let us first consider the writer process. As by assumption it is correct, it 
broadcasts the message {\sc write}$(sn,-)$  (line~\ref{F-SWMR-01}). 
Each correct process broadcasts {\sc write}$(sn,-)$ when it receives it 
for the first time (line~\ref{F-SWMR-08}). 
As there are at least $(n-t)$ correct processes, the writer eventually 
receives  {\sc write}$(sn,-)$ from these processes, and stops
waiting at line~\ref{F-SWMR-02}.  

Let us now consider a correct reader process $p_i$. It follows from 
the  same reasoning as before that the reader receives 
the message {\sc state}$(rsn,-)$ from at least $(n-t)$ processes
(lines~\ref{F-SWMR-05} and~\ref{F-SWMR-12}).   Hence, it remains to 
prove that the second part of the waiting predicate, namely 
$swsn_i\geq maxwsn$ (line~\ref{F-SWMR-05}) becomes eventually true, 
where $maxwsn$ is the greatest write sequence number received by $p_i$ in a 
message {\sc state}$(rsn,-)$. Let $p_j$ be the sender of this message. 
The following list of items is such that item $x$ $\implies$ item $(x+1)$, 
from which follows that $swsn_i\geq maxwsn$ (line~\ref{F-SWMR-05}) is 
eventually satisfied.
\begin{enumerate}
\vspace{-0.1cm}
\item 
$p_j$ updated $wsn_j$ to $maxwsn$ (line~\ref{F-SWMR-07})
before sending  {\sc state}$(rsn,maxwsn)$  (line~\ref{F-SWMR-12}). 
\vspace{-0.2cm}
\item 
Hence, $p_j$ received previously the message  {\sc write}$(maxwsn,-)$, and
broadcast it the first time it  received it (line~\ref{F-SWMR-08}).
\vspace{-0.2cm}
\item 
It follows that any correct process receives the message {\sc write}$(maxwsn,-)$
(at least from $p_j$), and broadcasts it the first time it receives it
(line~\ref{F-SWMR-08}). 
\vspace{-0.2cm}
\item 
Consequently, $p_i$ eventually receives the message  {\sc write}$(maxwsn,-)$ 
from $(n-t)$ processes. When this occurs, it updates $swsn_i$
(line~\ref{F-SWMR-10}), which is then $\geq maxwsn$, which concludes 
the proof of the termination of a read operation.
\end{enumerate} 
\vspace{-0.4cm}
\renewcommand{\toto}{lemma:termination}
\end{proofL}

\begin{lemma}
\label{lemma:atomicity}
The register $\REG$ is atomic. 
\end{lemma}

\begin{proofL}
Let $read[i,x]$ be a read operation issued by a process $p_i$ which returns 
the value with sequence number $x$, and $write[y]$ be the write operation
which writes the value with sequence number $y$. 
The proof of the  lemma is the consequence of the three following claims. 
\begin{itemize}
\vspace{-0.1cm}
\item Claim 1. 
If $read[i,x]$ terminates before $write[y]$ starts, then $x<y$.
\vspace{-0.2cm}
\item Claim 2. 
If $write[x]$ terminates before $read[i,y]$ starts, then $x\leq y$.
\vspace{-0.2cm}
\item Claim 3. 
If $read[i,x]$ terminates before $read[j,y]$ starts, then $x\leq y$.
\end{itemize}
Claim 1 states that no process can read from the future. 
Claim 2 states that no process can read overwritten values.
Claim 3 states that there is no new/old read inversions~\cite{AW04,R13-a}. \\

\noindent
Proof of Claim 1. \\
This claim follows from the following simple observation. When the writer 
executes $write[y]$, it first increases its local variable $wsn$ which 
becomes greater than any  sequence number associated with its previous 
write operations (line~\ref{F-SWMR-01}). Hence if  $read[i,x]$ terminates 
before $write[y]$ starts, we necessarily have $x<y$.\\

\noindent
Proof of Claim 2. \\
It follows from line~\ref{F-SWMR-02} and lines~\ref{F-SWMR-07}-\ref{F-SWMR-08} 
that, when $write[x]$ terminates, there is a set $Q_w$ of  at least $(n-t)$ 
processes $p_k$ such that $wsn_k\geq x$. On another side, 
due to lines~\ref{F-SWMR-04}-\ref{F-SWMR-05} and 
line~\ref{F-SWMR-12},  $read[i,y]$ obtains a message {\sc state}$()$
from a set $Q_r$ of  at least $(n-t)$ processes.

As $|Q_w|\geq n-t$, $|Q_r|\geq n-t$, and $n>2t$, it follows that
$Q_w \cap Q_r$ is not empty. There is consequently a process 
$p_k\in Q_w \cap Q_r$, such that that $wsn_k\geq x$. 
Hence, $p_k$  sent to $p_i$ the message  {\sc state}$(-,z)$, where $z\geq x$.

 Due to (a) the definition of $maxwsn\geq z$, 
(b) the predicate $swsn_i\geq  maxwsn \geq   z$ (line~\ref{F-SWMR-05}), 
and (c) the value of  $swsn_i=y$, 
it follows that $y= swsn_i\geq z$ when $read[i,y]$ stops waiting at
line~\ref{F-SWMR-05}. As, $z\geq x$, it follows $y\geq x$,
which proves the claim. ~\\

\noindent
Proof of Claim 3. \\
When $read[i,x]$ stops waiting at line~\ref{F-SWMR-05},
it returns the value $res_i$ associated with the sequence number $swsn_i=x$.
Process $p_i$ previously  received the message {\sc write}$(x,-)$
from a set $Q_{r1}$ of at least $(n-t)$ processes. 
The same occurs for $p_j$, which, before returning, received the message 
{\sc write}$(y,-)$ from a set $Q_{r2}$ of at least $(n-t)$ processes.

As $|Q_{r1}|\geq n-t$, $|Q_{r2}|\geq n-t$, and $n>2t$, it follows that
$Q_{r1} \cap Q_{r2}$ is not empty. Hence, there is a process $p_k$
which sent {\sc state}$(,x)$ to $p_i$,
and later sent  {\sc state}$(-,y)$ to $p_j$. 
As $swsn_k$ never decreases, it follows that $x\leq y$, which completes 
the proof of the lemma. 
\renewcommand{\toto}{lemma:atomicity}
\end{proofL}

\begin{theorem}
\label{theo:main-1}
Algorithm {\em\ref{algo:fast-basic-version}} implements an {\em SWMR}
atomic register in $\CAMP_{n,t}[t<n/2]$. 
\end{theorem}

\begin{proofT}
The proof follows from Lemma~\ref{lemma:termination} (termination) 
and Lemma~\ref{lemma:atomicity} (atomicity).
\renewcommand{\toto}{theo:main-1}
\end{proofT}

\subsection{Time-efficiency: the {\it bounded delay} assumption}
As already indicated, this underlying synchrony assumption 
considers that every message takes at most $\Delta$ time units.
Moreover,  let us remind that  a read
(which started at time $\tau_r$) is write-latency-free if it is not
concurrent with a write, and the last preceding write did not crash and
started at time $\tau_w< \tau_r-\Delta$.

\begin{lemma}
\label{lemma:fast-writer}
A write operation takes at most $2\Delta$ time units.
\end{lemma}

\begin{proofL}
The case of the writer is trivial. The message {\sc write}$()$
broadcast by the writer takes at most $\Delta$ time units, as do the 
acknowledgment messages{\sc write}$()$ sent by each process to the writer. 
In this case $2\Delta$ correspond to a causality-related
maximal round-trip delay (the reception of a message triggers the
sending of an associated acknowledgment).
\renewcommand{\toto}{lemma:fast-writer}
\end{proofL}

\paragraph{When the writer does not crash while executing a write operation}
The cases where the writer  does not crash while executing a write operation 
are captured by the next two lemmas.

\begin{lemma}
\label{lemma:fast-read-write-correct-writer-2}
A write-latency-free read operation takes at most $2\Delta$ time units.
\end{lemma}

\begin{proofL}
Let $p_i$ be a process that issues a write-latency-free read operation, 
and  $\tau_r$ be its starting time. Moreover, 
Let $\tau_w$ the starting time of the last preceding write. 
As the read is write latency-free, we have $\tau_w+\Delta <\tau_r$. 
Moreover, as messages take at most $\Delta$ time units, 
and the writer did not crash when executing the write, 
each non-crashed process $p_k$ received the message {\sc write}$(x,-)$ 
(sent by the preceding write at time $\tau_w+\Delta < \tau_r$), broadcast it
(line~\ref{F-SWMR-08}), and updated its local variables  such that 
we have $wsn_k=x$ (lines~\ref{F-SWMR-07}-\ref{F-SWMR-11}) at 
ime $\tau_w+\Delta < \tau_r$. Hence, all the messages {\sc state}$()$ 
received by the reader $p_i$  carry the write sequence number $x$.
Moreover, due to the broadcast of 
line~\ref{F-SWMR-08} executed by each correct process, we have 
$swsn_i=x$ at some time $\tau_w+2\Delta < \tau_r+\Delta$.
It follows that the predicate of line~\ref{F-SWMR-05} 
is satisfied at $p_i$ within $2\Delta$ time units after it invoked the 
read operation.
\renewcommand{\toto}{lemma:fast-read-write-correct-writer-2}
\end{proofL}
\begin{lemma}
\label{lemma:fast-read-write-correct-writer-3}
A read operation which is not  write-latency-free, and during which 
the writer does not crash during the interfering write operation,  
takes at most $3\Delta$.
\end{lemma}

\begin{proofL}
Let us consider a  read operation that starts at time $\tau_r$, 
concurrent with a  write operation that starts at time 
$\tau_w$ and during which the writer does not crash.  
From the  read operation point of view, the worst case occurs when the read 
operation is invoked just after time $\tau_w-\Delta$, let us say at 
time $\tau_r=\tau_w-\Delta+\epsilon$. 
As a message {\sc state}$(rsn,-)$ is sent by return when a message
{\sc read}$(rsn)$ is received, the messages {\sc state}$(rsn,-)$
received by $p_i$ by time $\tau_r+2\Delta$ can be such that some carry
the sequence number $x$ (due to last previous write) while others
carry the sequence number $x+1$ (due to the concurrent write)\footnote{
Messages {\sc state}$(rsn,x)$ are sent by the  processes
that received  {\sc read}$(rsn)$ before $\tau_w$, while 
the messages {\sc state}$(rsn,x+1)$ are sent by the  processes
that received  {\sc read}$(rsn)$ between $\tau_w$ and  
$\tau_r+\Delta= \tau_w+\epsilon$.}.  Hence,
$maxwsn=x$ or $maxwsn=x+1$ (predicate of line~\ref{F-SWMR-05}).  If
$maxwsn=x$, we also have $swsn_i=x$ and $p_i$ terminates its read.  If
$maxwsn=x+1$, $p_i$ must wait until $swsn_i=x+1$, which occurs at the
latest at $\tau_w+2\Delta$ (when $p_i$ receives the last message of
the $(n-t)$ messages {\sc write}$(y,-)$ which makes true the predicates
of lines~\ref{F-SWMR-09}-\ref{F-SWMR-10}, thereby  allowing the predicate 
of line~\ref{F-SWMR-05} to be satisfied).
When this occurs, $p_i$ terminates its read operation.  
As $\tau_w=\tau_r+\Delta-\epsilon$, $p_i$ returns at the latest
$\tau_r+3\Delta-\epsilon$ time units after it invoked the read operation.  
\renewcommand{\toto}{lemma:fast-read-write-correct-writer-2}
\end{proofL}

\paragraph{When the writer crashes while executing a write operation}
The problem raised by the crash of the writer while executing 
the write operation is when it crashes while  broadcasting 
the message {\sc write}$(x,-)$ (line~\ref{F-SWMR-01}): some processes
receive this message by $\Delta$ time units, while other processes do not. 
This issue is solved by the propagation of the  message {\sc write}$(x,-)$
by the non-crashed processes that receive it  (line~\ref{F-SWMR-08}).
This means that, in the worst case (as in synchronous systems), 
the message {\sc write}$(x,-)$ must  be forwarded by $(t+1)$ processes 
before being received by all correct processes. This worst scenario may 
entail a cost of $(t+1)\Delta$ time units. 

\begin{figure}[th]
\centering{\fbox{
\begin{minipage}[t]{150mm}
\footnotesize
\renewcommand{\baselinestretch}{2.5}
\resetline
\begin{tabbing}
aaaaa\=aaa\=aaaaa\=aaaaaa\=\kill

{\bf when} {\sc write}($wsn,v$) 
     \underline{or {\sc state}$(rsn,wsn,v)$} {\bf is received do}\\

(\ref{F-SWMR-07})
\> {\bf if} ($wsn>wsn_i$) 
   {\bf then}  $reg_i \leftarrow v$; $wsn_i\gets wsn$;  
               $\mathsf{broadcast}$ {\sc write}($wsn,v$)
{\bf end if}; \\

(\ref{F-SWMR-08}) 
\> {\bf if} (not yet done)  {\bf then}   
    $\mathsf{broadcast}$ {\sc write}($wsn,v$)
{\bf end if}; \\

(\ref{F-SWMR-09}) 
\> {\bf if} \big({\sc write}($wsn,-$)  received 
                 from $(n-t)$ different processes\big)  \\

(\ref{F-SWMR-10})   
\> \> {\bf then} 
   {\bf if} $(wsn > swsn_i)$ $\wedge$ (not already done)  {\bf then}
             $swsn_i\gets wsn$;  $res_i\gets v $ {\bf end if}\\

(\ref{F-SWMR-11}) 
\>   {\bf end if}. \\~\\

{\bf when} 
{\sc read}$(rsn)$ {\bf is received from} $p_j$ {\bf do}\\

(\ref{F-SWMR-12}) 
\> $\mathsf{send}$ 
            {\sc state}$(rsn,wsn_i\underline{,reg_i})$ $\mathsf{to}$ $p_j$. 

\end{tabbing}
\end{minipage}
}
\caption{Modified algorithm for time-efficient read in case of 
concurrent writer crash}
\label{algo-modified}
}
\end{figure}

Figure~\ref{algo-modified} presents a simple modification of 
Algorithm~\ref{algo:fast-basic-version}, which allows a fast implementation 
of read operations whose executions are concurrent with a write operation 
during which the writer crashes.  The modifications are underlined. 

When a process $p_i$  receives a message {\sc read}$()$, it now returns a
message {\sc state}$()$  containing an additional field, namely the 
current value of $reg_i$, its local copy of $\REG$  (line~\ref{F-SWMR-12}).

When a process $p_i$ receives from a process $p_j$  a message 
{\sc state}$(-,wsn,v)$,  it uses it in the waiting predicate  
of line~\ref{F-SWMR-05}, 
but executes before the lines~\ref{F-SWMR-07}-\ref{F-SWMR-11}, as if this 
message was  {\sc write}$(wsn,v)$.  
According to the values of the predicates of lines~\ref{F-SWMR-07}, 
~\ref{F-SWMR-09}, and~\ref{F-SWMR-10}, this allows  $p_i$ to expedite 
the update of its local variables $wsn_i$, $reg_i$, $swsn_i$, and $res_i$, 
thereby favoring  fast termination.

The reader can check that these modifications do not alter the proofs 
of Lemma~\ref{lemma:termination} (termination) 
and Lemma~\ref{lemma:atomicity} (atomicity). 
Hence, the proof of Theorem~\ref{theo:main-1} is still correct.

\begin{lemma}
\label{lemma:faulty-writer}
A read operation which is not  write-latency-free, and during which the 
writer crashes during the interfering write operation,  takes at most 
$4\Delta$ time units.
\end{lemma}

\begin{proofL}
Let $\tau_r$ be  the time at which the read operation starts.  
As in the proof of Lemma~\ref{lemma:fast-read-write-correct-writer-2},
the messages {\sc state}$(rsn,-,-)$
received $p_i$ by time $\tau_r+2\Delta$ can be such that some carry
the sequence number $wsn=x$ (due to last previous write) while some others
carry the sequence number $wsn=x+1$ (due to the concurrent write during 
which the writer crashes). If all these messages carry $wsn=x$, 
the read terminates by time  $\tau_r+2\Delta$. If at least one of these 
messages is {\sc state}$(rsn,x+1,-)$, we have $maxwsn = x+1$, and  $p_i$ waits 
until the predicate $swsn_i\geq maxwsn~(=x+1)$ becomes true
(line~\ref{F-SWMR-05}).

When it received  {\sc state}$(rsn,x+1,-)$, if not yet done, 
$p_i$ broadcast  the message 
{\sc write}$(rsn,x+1,-)$, (line~\ref{F-SWMR-08} of Figure~\ref{algo-modified}), 
which is received by the other processes within $\Delta$ time units. 
If not yet done, this entails the broadcast by each correct process of 
the same message  {\sc write}$(rsn,x+1,-)$. Hence, at most $\Delta$ time 
units later,  
$p_i$ has received the message {\sc write}$(rsn,x+1)$ from $(n-t)$ processes, 
which entails the update of $swsn_i$ to $(x+1)$. Consequently the 
predicate of line~\ref{F-SWMR-05} becomes satisfied, and $p_i$ terminates 
its read operation. 

 When counting the number  of consecutive communication steps, we have:
The message {\sc read}$(rsn)$ by $p_i$, followed by a message 
{\sc state}$(rsn,x+1,-)$ sent by some process and received by $p_i$,   
followed by the message {\sc write}$(rsn,x+1)$ broadcast by $p_i$, 
followed by  the message {\sc write}$(rsn,x+1)$ broadcast by each 
non-crashed process (if not yet done).   
Hence, when the writer crashes during a concurrent read, 
the read returns within at most $\tau_r+4\Delta$ time units. 
\renewcommand{\toto}{lemma:faulty-writer}
\end{proofL}

\begin{theorem}
\label{theo:main-2}
Algorithm {\em\ref{algo:fast-basic-version}} modified as indicated in
Figure~{\em\ref{algo-modified}} implements in $\CAMP_{n,t}[t<n/2]$ an {\em SWMR}
 atomic register with time-efficient operations  (where the time-efficiency 
notion is based on the {\em bounded delay} assumption). 
\end{theorem}

\begin{proofT}
The proof follows from Theorem~\ref{theo:main-1} (termination and atomicity),
Lemma~\ref{lemma:fast-writer},
Lemma~\ref{lemma:fast-read-write-correct-writer-2}, 
Lemma~\ref{lemma:fast-read-write-correct-writer-3}, and 
Lemma~\ref{lemma:faulty-writer} (time-efficiency). 
\renewcommand{\toto}{theo:main-2}
\end{proofT}

\subsection{Time-efficient implementation: 
the {\it round-based synchrony} assumption}
As already indicated, this notion of a time-efficient implementation assumes 
an underlying round-based synchronous communication system, where the
duration of a round (duration of all message transfer delays) is $\delta$.

\begin{lemma}
\label{lemma:writer}
The duration of write operation is $2\delta$.  
\end{lemma}

\begin{proofL}
The proof follows directly from the observation that the write operation 
terminates  after a round-trip delay, whose duration is $2\delta$.  
\renewcommand{\toto}{lemma:writer}
\end{proofL}

\begin{lemma}
\label{lemma:reader}
The duration of a read operation is $2\delta$ time units
if the writer does not crash while executing a write operation concurrent 
with the read. Otherwise, it can be $3\delta$. 
\end{lemma}

\begin{proofL}
Considering a read operation that starts at time $\tau_r$, let us assume that 
the writer does not crash while concurrently executing a write operation. 
At time $\tau_r+\delta$ all processes receives the message 
{\sc read}$(rsn)$ sent by the reader (line~\ref{F-SWMR-04}), 
and answer with a  message {\sc state}$(rsn,-)$  (line~\ref{F-SWMR-12}).
Due the round-based synchrony assumption, all these messages carry the same 
sequence number $x$, which is equal to both their local variable 
$wsn_i$ and $swsn_i$. It follows that at time  $\tau_r+2\delta$, the
predicate of line~\ref{F-SWMR-05} is satisfied at the reader, which 
consequently returns from the read operation. 

If the writer  crashes while concurrently executing a write operation, 
it is possible that during some time (a round duration), 
some processes know the sequence number $x$, while other processes know
only $x-1$. But this synchrony break in the knowledge of the last sequence 
number is  mended during the next round thanks to the message 
{\sc write}$(x,v)$ sent by the processes which are aware of $x$
(See Figure~\ref{algo-modified}). 
After this additional round, the read terminates (as previously) in two 
rounds. Hence, the read returns at the latest at time $\tau_r+3\delta$.
\renewcommand{\toto}{lemma:reader}
\end{proofL}

\begin{theorem}
\label{theo:main-3}
Algorithm {\em\ref{algo:fast-basic-version}} modified as indicated in
Figure~{\em\ref{algo-modified}} implements in $\CAMP_{n,t}[t<n/2]$ an 
{\em SWMR} atomic register with time-efficient operations (where the 
time-efficiency notion is based on the {\em round-based synchrony} assumption). 
\end{theorem}

\begin{proofT}
The proof follows from Theorem~\ref{theo:main-1} (termination and atomicity),
Lemma~\ref{lemma:writer}, and Lemma~\ref{lemma:reader} (time-efficiency).
\renewcommand{\toto}{theo:main-3}
\end{proofT}

\section{Conclusion}
\label{sec:conclusion}
This work has presented a new distributed algorithm implementing an
atomic read/write register on top of an  asynchronous $n$-process 
message-passing system in which up to $t<n/2$ processes may crash.
When designing it, the constraints we imposed on this algorithm were 
(a) from an efficiency point of view: provide  time-efficient implementations 
for read and write operations, (b) and from a design principle point of view: 
remain ``as close as possible'' to the flagship ABD algorithm introduced by
Attiya, Bar-Noy and Dolev~\cite{ABD95}.\\

The ``time-efficiency'' property of the proposed algorithm has been 
analyzed  according to two synchrony assumptions on the underlying system.  
\begin{itemize}
\vspace{-0.1cm}
\item
The first assumption considers  an upper bound $\Delta$ 
on message transfer delays. Under such an assumption, 
any write operation takes then  at most $2\Delta$ time units, 
and a read operation  takes at most $2\Delta$ time units when executed 
in good circumstances (i.e., when there is no write operation concurrent 
with the read operation).  
Hence, the inherent cost of an operation is a round-trip delay, 
always for a write and in favorable circumstances for a read. 
A read operation concurrent with a write operation during which the writer 
does not crash,  may require an additional cost of $\Delta$, which means 
that it takes at most $3\Delta$ time units.
Finally, if the writer crashes during a write concurrent with a  read,
the read may take at most $4\Delta$ time units.
This shows clearly the incremental cost imposed by the adversaries
(concurrency of write operations, and failure of the writer). 
\vspace{-0.2cm}
\item
The second assumption investigated for a ``time-efficient implementation'' is
the one provided by  a round-based synchronous system, where message transfer
delays (denoted $\delta$) are assumed to be the same for all messages.
It has been shown that, under this assumption, the duration of a write
is $2\delta$, and the duration of a read is $2\delta$, or
exceptionally $3\delta$ when the writer crashes while concurrently
executing a write operation.
\end{itemize}

It is important to remind that the proposed algorithm remains correct 
in the presence of any asynchrony pattern. Its time-efficiency features
are particularly interesting when the system has long synchrony periods.

Differently from the proposed algorithm, the ABD algorithm does not 
display different behaviors in different concurrency  and failure patterns.  
In ABD, the duration of all  write operations is
upper bounded by  $2\Delta$ time units (or equal to $2\delta$), and  
the duration of all  read operations is upper bounded by 
$4\Delta$ time units (or equal to $4\delta$).
The trade-off  between ABD and our algorithm lies 
the message complexity,  which is $O(n)$ in ABD for both read and write 
operations, while it is  $O(n^2)$ for a write operation and  $O(n)$ 
for a read  operation in the proposed algorithm. Hence our algorithm is 
particularly interesting for registers used in read-dominated applications. 
Moreover, it helps us better understand the impact of the adversary pair 
``writer concurrency + writer failure'' on the efficiency of the read 
operations.

\section*{Acknowledgments}
This work has been partially supported by the  
Franco-German DFG-ANR Project 40300781 DISCMAT
devoted to connections between mathematics and distributed computing, and 
the French  ANR project DISPLEXITY devoted to the study of computability 
and complexity  in distributed computing. 



\end{document}